\documentstyle[aps,prd,preprint,eqsecnum]{revtex}
\begin{document}

\title{There is no new physics in the multiplicative anomaly}

\author{J.~J.~McKenzie-Smith\thanks{email: 
{\tt j.j.mckenzie-smith@ncl.ac.uk}} and D.~J.~Toms\thanks{email: {\tt d.j.toms@ncl.ac.uk}}}

\address{Department of Physics, University of Newcastle upon Tyne,\\ Newcastle upon Tyne, NE1~7RU, U.~K.}

\date{May 1998}
\maketitle
\begin{abstract}
We discuss the role of the multiplicative anomaly for a complex scalar
field at finite temperature and density. It is argued that physical considerations must be applied to determine which of the many possible expressions for the effective action obtained by the functional integral method is correct. This is done by first studying the non-relativistic field where the thermodynamic potential is well-known. The relativistic case is also considered. We emphasize that the role of the multiplicative anomaly is not to lead to new physics, but
rather to preserve the equality among the various expressions for the effective action.
\end{abstract}

\section{Introduction}
\label{sec1}

The Feynman functional integral (or path integral) is now the most widely used approach to quantum field theory at zero as well as finite temperature. Of course, it is perfectly possible to use more standard techniques based on operator methods with the Feynman path integral never appearing. What must be the case is that regardless of the approach adopted, all valid methods lead to the same physical consequences. Because of the lack of rigour in defining the Feynman functional integral, care must be exercised in accepting the results of the formal manipulations involved. In the end we can only accept the results of the functional integral approach if they are in agreement with other methods of calculation.

In a one-loop calculation of the effective action for a scalar field theory the result typically involves the determinant of a differential operator, a result which is infinite and must be defined through some regularisation technique. If we use 
the background field method \cite{DeWitt}, then the one-loop effective action reads
\begin{equation}
\Gamma =S[\bar{\phi}] + \frac{1}{2}\ln\det\left(\ell^{2}S_{,ij}[\bar{\phi}]\right)
\label{AA}
\end{equation}
where $\bar{\phi}$ denotes the background field, $S[\bar{\phi}]$ is the classical action functional, and
\begin{equation}
S_{,ij}[\bar{\phi}] =\frac{\delta^{2}S[\bar{\phi}]}{\delta\bar{\phi}^{i}(x)
\delta\bar{\phi}^{j}(x')}\;.
\label{AB}
\end{equation}
The second term of (\ref{AA}) contains the quantum corrections to the classical theory (at one-loop order) and arises from performing the functional integral over a gaussian. $\ell$ in (\ref{AA}) is a unit of length (the renormalization scale) introduced in order that the argument of the logarithm in (\ref{AA}) be dimensionless. In finite temperature field theory it is convenient to adopt the imaginary time formalism in which the path integral extends over all fields periodic in time with period $\beta=1/T$ with $T$ the temperature. We choose to work with real scalar fields.

The problem of evaluating the second term of (\ref{AA}) now arises. To be explicit we will consider the case of a single complex scalar field at finite charge density. Writing the complex field in terms of its real and imaginary parts we have \cite{Kap,HW,BBD,DJT,KKDJT}
\begin{eqnarray}
S[\bar{\phi}] & = &\int_{0}^{\beta}dt\int_{\Sigma}d\sigma_{x}
\left(\frac{1}{2}(\dot{\bar{\phi_{1}}} - ie\mu\bar{\phi_{2}})^{2} + 
\frac{1}{2}(\dot{\bar{\phi_{2}}} + ie\mu\bar{\phi_{1}})^{2}\right.  \nonumber \\
 & & \left. + \frac{1}{2}|\nabla\bar{\phi_{1}}|^{2} + \frac{1}{2}|\nabla\bar{\phi_{2}}|^{2} + \frac{1}{2}m^{2}(\bar{\phi_{1}^{2}}+\bar{\phi_{2}^{2}}) + \frac{\lambda}{4!}(\bar{\phi_{1}^{2}}+\bar{\phi_{2}^{2}})^{2}\right)\;.
\label{AC}
\end{eqnarray}
Here $\Sigma$ is the spatial part of the spacetime, $e$ is the electronic charge and $\mu$ is the chemical potential. If we choose $\bar{\phi_{1}}=\bar{\phi}$ and $\bar{\phi_{2}}=0$ as the background fields it is easy to show that
\begin{equation}
\frac{\delta^{2}S[\bar{\phi}]}{\delta\bar{\phi_{1}}(x)
\delta\bar{\phi_{1}}(x')} = \left( - \Box_{x} + m^{2} - e^{2}\mu^{2} + \frac{\lambda}{2}\bar{\phi}^{2}\right)\delta(x,x')\;,
\label{AD}
\end{equation}
\begin{equation}
\frac{\delta^{2}S[\bar{\phi}]}{\delta\bar{\phi_{2}}(x)
\delta\bar{\phi_{2}}(x')} = \left( - \Box_{x} + m^{2} - 
e^{2}\mu^{2} + \frac{\lambda}{6}\bar{\phi}^{2}\right)\delta(x,x')\;,
\label{AE}
\end{equation}
\begin{equation}
\frac{\delta^{2}S[\bar{\phi}]}{\delta\bar{\phi_{1}}(x)
\delta\bar{\phi_{2}}(x')}
= - \frac{\delta^{2}S[\bar{\phi}]}{\delta\bar{\phi_{2}}(x)
\delta\bar{\phi_{1}}(x')} = 2ie\mu\frac{\partial}{\partial t}
\delta(x,x')\;.
\label{AF}
\end{equation}
There are now several approaches one could take to evaluate $\ln\det(\ell^{2}S_{,ij}[\bar{\phi}])$. To illustrate this we will simplify the problem by taking $\lambda=0$, so that the theory is free. We will comment on the interacting case at the end of our paper.

The first way is to take the determinant over the $2\times 2$ matrix first to obtain
\begin{eqnarray}
\Gamma^{(1)}_{A} & = & \frac{1}{2}\ln\det\ell^{2}\left(l^{2}S_{,ij}[\bar{\phi}]\right)
\nonumber \\
 & = & \frac{1}{2}\ln\det\ell^{4}\left(( - \Box_{x} + m^{2} - e^{2}\mu^{2})^{2} - 4e^{2}\mu^{2}
\frac{\partial^{2}}{\partial t^{2}}\right)\;.
\label{AG}
\end{eqnarray}
The remaining determinant in this last expression is understood to be a functional one. At this stage we note that this step has been criticised recently by Dowker \cite{Dowk} who claims that it is incorrect to take the $2\times 2$ determinant first. As we will show in the next section, this need not be an incorrect step. The relativistic case will be considered in Sec.~\ref{sec3}.

A second expression for $\Gamma^{(1)}$ is obtained by diagonalizing the functional matrix $S_{,ij}[\bar{\phi}]$ by a transformation of the fields to give
\begin{eqnarray}
\Gamma_{B}^{(1)} & = & -\frac{1}{2}\ln\det \ell^{2}\left[-\Box_{x} + m^{2} -e^{2}\mu^{2} + 2ie\mu\frac{\partial}{\partial t}\right] \nonumber \\
 &&\quad -\frac{1}{2}\ln\det \ell^{2}\left[-\Box_{x} + m^{2} -e^{2}\mu^{2} - 2ie\mu\frac{\partial}{\partial t}\right]\;.
\label{CI}
\end{eqnarray}
This can also be understood to correspond to a particular factorization of (\ref{AG}).

A third way to evaluate the effective action is to integrate over one of the fields, say $\phi_1$, first, and then perform the remaining integral over $\phi_2$. This leads to a result
\begin{eqnarray}
\Gamma_{C}^{(1)} & = & -\frac{1}{2}\ln\det \ell^{2}(-\Box_{x} + m^{2} - e^{2}\mu^{2}) \nonumber \\
 & & \: -\frac{1}{2}\ln\det \ell^{2}\left[(-\Box_{x} + m^{2} - e^{2}\mu^{2})\rule{0mm}{6mm}\right. \nonumber \\
 & & \left.\: - 4e^{2}\mu^{2}\frac{\partial^{2}}{\partial t^{2}}(-\Box_{x} + m^{2} - e^{2}\mu^{2})^{-1}\right]\;.
\label{CL}
\end{eqnarray}
The inverse operator $(-\Box_{x} + m^{2} -e^{2}\mu^{2})^{-1}$ can be understood to be the Green function. This expression is equivalent to factoring out $(-\Box_x+m^2-e^2\mu^2)$ from the determinant in (\ref{AG}).

At the formal level, by adopting the usual rules for the manipulation of finite dimensional matrices, all of the three expressions for the effective action are identical. However, as pointed out in Refs.~\cite{Eliz1,Eliz2}, the usual formal manipulations are vitiated by the presence of an anomaly, called the multiplicative anomaly. The essential feature is that this multiplicative anomaly leads to a difference between the three expressions for the effective action $\Gamma^{(1)}_{A,B,C}$. The question arises as to which, if any, of the three expressions is correct. Our viewpoint is that this issue cannot be settled by considering the formal Feynman functional integral without recourse to physics. We will justify this in the subsequent sections. In Sec.~\ref{sec2} we will consider the simpler non-relativistic theory where there is no question as to the correct thermodynamics. In Sec.~\ref{sec3} we will return to the relativistic field.

\section{Non-relativistic Theory}
\label{sec2}

The action functional for a non-relativistic Schr\"{o}dinger field $\Psi$ is (in the imaginary time formalism)
\begin{equation}
S[\Psi,\Psi^{\dagger}] = \int_{0}^{\beta}dt\int_{\Sigma}d\sigma_{x}\left(\frac{1}{2}
(\Psi^{\dagger}\frac{\partial\Psi}{\partial t} - \frac{\partial\Psi^{\dagger}}{\partial t}\Psi) + \frac{1}{2m}|\nabla\Psi|^{2} - \mu|\Psi|^{2}\right)\;.
\label{AH}
\end{equation}
We will not include self-interactions here for simplicity. The theory defined by (\ref{AH}) is already complicated enough to demonstrate the discrepancy between different approaches. Rather than deal with the complex field $\Psi$ we can decompose $\Psi$ 
into its real and imaginary parts and define
\begin{equation}
\Psi = \frac{1}{\sqrt{2}}(\phi_{1} + i\phi_{2})\;.
\label{AI}
\end{equation}
This gives
\begin{eqnarray}
S[\phi_{1},\phi_{2}] & = & \int_{0}^{\beta}dt\int_{\Sigma}d\sigma_{x}\left(\frac{i}{2}
(\phi_{1}\dot{\phi_{2}} - \phi_{2}\dot{\phi_{1}})\right. \nonumber \\
 & & \left.\: + \frac{1}{4m}|\nabla\phi_{1}|^{2} + \frac{1}{4m}|\nabla\phi_{2}|^{2} - \frac{1}{2}\mu(\phi_{1}^{2} + \phi_{2}^{2})\right)\;.
\label{AJ}
\end{eqnarray}
If we perform the Feynman functional integral over the two real fields in (\ref{AJ}) it is easy to see that the one-loop part of $\Gamma$ is
\begin{eqnarray}
\Gamma^{(1)} &=& \frac{1}{2}\ln\det \ell^{2}\left( \begin{array}{cc}
- \frac{1}{2m}\nabla^{2} - \mu & i\frac{\partial}{\partial t} \\
- i\frac{\partial}{\partial t} & \frac{1}{2m}\nabla^{2} - \mu
\end{array} \right)
\label{AK}\\
&=&\frac{1}{2}\ln\det \ell^{4}\left[\left(- \frac{1}{2m}\nabla^{2} - \mu\right)^{2} - \frac{\partial^{2}}{\partial t^{2}}\right]\;.
\label{AL}
\end{eqnarray}
In the second line above we have taken the determinant of the $2\times 2$ matrix which we claim is a valid step and that (\ref{AL}) is the correct answer. (This will be justified below).

In order to define (\ref{AL}) we make use of $\zeta$-function regularisation. Let $\{f_{n}(x)\}$ be a complete orthonormal set of solutions to
\begin{equation}
- \nabla^{2}f_{n}(x) = \sigma_{n}f_{n}(x)\;.
\label{AM}
\end{equation}
$\sigma_{n}$ represent the eigenvalues of the Laplacian for $f_{n}(x)$ obeying whatever boundary conditions apply to the space $\Sigma$. Because the fields $\phi_{1}$ and $\phi_{2}$ are periodic in imaginary time, the eigenvalues of the operator in (\ref
{AL}) are $(\sigma_{n}/2m - \mu)^{2} + \omega_{j}^{2}$ where
\begin{equation}
\omega_{j} = \frac{2\pi j}{\beta}
\label{AN}
\end{equation}
with $j = 0,\pm1,\pm2,\ldots$. We have the formal result
\begin{equation}
\Gamma^{(1)} = \frac{1}{2}\sum_{j,n}\ln
\ell^{4}\left[\left(\frac{\sigma_{n}}{2m} - \mu\right)^{2} + 
\omega_{j}^{2}\right]\;.
\label{AO}
\end{equation}
To give meaning to (\ref{AO}) we introduce the generalised $\zeta$-function \cite{Ray,Hawk,DowCr} defined by
\begin{equation}
\zeta(s) = \sum_{j=-\infty}^{+\infty}\sum_{n}\left[\left(\frac{\sigma_{n}}{2m} - \mu\right)^{2} + \omega_{j}^{2}\right]^{-s}
\label{AP}
\end{equation}
$\Gamma^{(1)}$ is then defined to be
\begin{equation}
\Gamma^{(1)} = -\frac{1}{2}\zeta'(0) + \frac{1}{2}\zeta(0)\ln \ell^{4}\;.
\label{AQ}
\end{equation}
Knowledge of $\zeta(s)$ in a neighbourhood of $s=0$ will therefore give us an expression for the one-loop effective action.

The $\zeta$-function is considered in the appendix. If we take $\mu=0$ in (\ref{App1}) and also $E_{n}=\sigma_{n}/2m - \mu$ then we may deduce $\zeta(0)$ and $\zeta'(0)$ from (\ref{App8}) once we have evaluated the energy $\zeta$-function of the
 first term.  From (\ref{App4}) we have
\begin{equation}
E(\alpha) = \sum_{n}\left(\frac{\sigma_{n}}{2m} - \mu\right)^{1-\alpha}
\label{AR}
\end{equation}
in this case. If we now specialise $\Sigma$ to be flat space with the large volume limit taken, then we have
\begin{equation}
\sigma_{n} \rightarrow k^{2}\;,
\label{AS}
\end{equation}
\begin{equation}
\sum_{n} \rightarrow V \int\frac{d^{3}k}{(2\pi)^{3}}\;.
\label{AT}
\end{equation}
It is now easy to show that
\begin{equation}
E(\alpha) = V \frac{\Gamma(\alpha - 5/2)}{\Gamma(\alpha - 1)}\left(\frac{m}{2\pi}\right)^{3/2}(-\mu)^{5/2 - \alpha}\;.
\label{AU}
\end{equation}
Assuming that $E(2s)$ is analytic at $s=0$, referring to (\ref{App8}) we find
\begin{eqnarray}
\zeta(0) &=& 0\;,
\label{AV}\\
\zeta'(0) &=&-\beta E(0) -2\sum_{n}\ln \left[1-e^{-\beta (\frac{\sigma_{n}}{2m}-\mu)}\right]\;.
\label{AW}
\end{eqnarray}
We therefore have, from (\ref{AQ}),
\begin{equation}
\Gamma^{(1)} =\frac{1}{2}\beta E(0)+ \sum_{n}\ln \left[1-e^{-\beta (\frac{\sigma_{n}}{2m}-\mu)}\right]\;.
\label{AX}
\end{equation}
The first term in $\Gamma^{(1)}$ is the regularized zero-point energy. In the simple case we are considering here, $E(0)=0$, so that the zero-point energy makes no contribution. We are left with the second term of (\ref{AX}) which is in agreement with what is written down in standard statistical mechanics. We can therefore be certain that (\ref{AL}) and $\zeta$-function regularisation have led to results which agree with those found by other methods. If we use (\ref{AS}) and (\ref{AT}) in (\ref{AX}) we find
\begin{equation}
\Gamma^{(1)} = -V \left(\frac{m}{2\pi\beta}\right)^{3/2}Li_{5/2}(e^{\beta\mu})
\label{AY}
\end{equation}
where
\begin{equation}
Li_{p}(z) = \sum_{n=1}^{\infty} \frac{z^{n}}{n^{p}}
\label{AZ}
\end{equation}
defines the polylogarithm.  This result may now be used to discuss Bose-Einstein condensation of the ideal gas. (See Ref.~\cite{Pathria} for example.)

In order to see what can go wrong, suppose that we return to (\ref{AL}). This time, by noting that
\begin{equation}
\left(-\frac{1}{2m}\nabla^{2} - \mu^{2}\right)^{2} - \frac{\partial^{2}}{\partial t^{2}} = \left(-\frac{1}{2m}\nabla^{2} - \mu - \frac{\partial}{\partial t}\right)\left(-\frac{1}{2m}\nabla^{2} - \mu + \frac{\partial}{\partial t}\right)
\end{equation}
we will write
\begin{equation}
\tilde{\Gamma}^{(1)}= \frac{1}{2}\ln\det \ell^{2}\left(-\frac{1}{2m}\nabla^{2} - \mu - \frac{\partial}{\partial t}\right)+ \frac{1}{2}\ln\det \ell^{2}\left(-\frac{1}{2m}\nabla^{2} - \mu + \frac{\partial}{\partial t}\right)\;.
\label{BA}
\end{equation}
Formally (\ref{BA}) and (\ref{AL}) are the same, since we have simply written the determinant of a product as a product of determinants. However, as pointed out in Refs.~\cite{Eliz1,Eliz2} this may not be justified for differential operators. Because of the simplicity of the model, we can evaluate $\tilde{\Gamma}^{(1)}$ in (\ref{BA}) explicitly and see if it agrees with $\Gamma^{(1)}$, which we know to be correct.

We will define the generalised $\zeta$-function
\begin{equation}
\tilde{\zeta}(s) = \sum_{j=-\infty}^{+\infty} \sum_{n}\left(\frac{\sigma_{n}}{2m} - \mu + i\omega_{j}\right)^{-s}\;.
\label{BB}
\end{equation}
Because the only difference between the two terms in (\ref{BA}) lies in the sign of $t$, and the sum over $j$ in (\ref{BB}) goes from $-\infty$ to $+\infty$ it is easy to see that
\begin{equation}
\tilde{\Gamma}^{(1)} = - \tilde{\zeta}'(0) + 
\tilde{\zeta}(0)\ln\ell^{2}\;.
\label{BC}
\end{equation}
This is also the result which would be obtained by performing the Feynman functional integral over the complex field $\Psi$ using (\ref{AH}). (See \cite{DJTPRB}).

From (\ref{AS}) and (\ref{AT}) we have
\begin{equation}
\tilde{\zeta}(s) = V \frac{\Gamma(s-3/2)}{\Gamma(s)}\left(\frac{m}{2\pi}\right)^{3/2} \sum_{j=-\infty}^{+\infty}(i\omega_{j} - \mu)^{3/2 - s}
\label{BD}
\end{equation}
after performing the integral over $k$. The sum in (\ref{BD}) was evaluated in the appendix of \cite{DJTPRB}. Making use of this result we find
\begin{equation}
\tilde{\zeta}(s) = V \frac{\Gamma(s-3/2)}{\Gamma(s)}
\left(\frac{m}{2\pi}\right)^{3/2}(-\mu)^{3/2-s} + 
V\left(\frac{m}{2\pi\beta}\right)^{3/2}
\frac{\beta^{s}}{\Gamma(s)}Li_{5/2 - s}(e^{\beta \mu})\;.
\label{BE}
\end{equation}
Expansion about $s=0$ shows that
\begin{equation}
\tilde{\zeta}(0) = 0\;,
\label{BF}
\end{equation}
\begin{equation}
\tilde{\zeta}'(0) = \frac{4}{3}\sqrt{\pi}V\left(\frac{m}{2\pi}\right)^{3/2}(-\mu)^{3/2} + V\left(\frac{m}{2\pi\beta}\right)^{3/2}Li_{5/2}(e^{\beta \mu})\;.
\label{BG}
\end{equation}
We therefore find
\begin{equation}
\tilde{\Gamma}^{(1)} = - \frac{4}{3}\sqrt{\pi}V\left(-\frac{m\mu}{2\pi}\right)^{3/2} - V\left(\frac{m}{2\pi\beta}\right)^{3/2}Li_{5/2}(e^{\beta \mu})
\label{BH}
\end{equation}
with the definition (\ref{BA}).

Comparison of (\ref{BH}) with the correct result (\ref{AY}) shows that
\begin{equation}
\Gamma^{(1)} - \tilde{\Gamma}^{(1)} = \frac{4}{3} \sqrt{\pi}V\left(-\frac{m\mu}{2\pi}\right)^{3/2}\;.
\label{BI}
\end{equation}
The results are different. It might be thought that the difference (\ref{BI}) is just a trivial constant which can lead to no physical consequences; however this is incorrect. The difference depends on the chemical potential $\mu$ and can affect thermodynamic quantities, such as the particle number or internal energy. Use of $\tilde{\Gamma}^{(1)}$ rather than $\Gamma^{(1)}$ will lead to results which are not in agreement with those found from statistical mechanics, and therefore not in agreement with observations of physical systems. In particular, the particle number follows from the normal Bose-Einstein distribution function as
\begin{eqnarray}
N &=& \sum_{n}\left[e^{\beta (\frac{\sigma_{n}}{2m}
- \mu)} - 1\right]^{-1}
\label{BJ}\\
&=& V\left(\frac{m}{2\pi\beta}\right)^{3/2}Li_{3/2}(e^{\beta\mu})\;.
\label{BK}
\end{eqnarray}
We also must have (since $\Gamma$ is related to the Helmholtz free energy)
\begin{equation}
N = -\frac{1}{\beta}\left(
\frac{\partial\Gamma^{(1)}}{\partial\mu}\right)_{\beta,V}\;.
\label{BL}
\end{equation}
The result (\ref{AY}) used in (\ref{BL}) agrees with (\ref{BK}), but if we use the result (\ref{BH}) in (\ref{BL}) we do not get the correct particle number. The only conclusion to be drawn is that (\ref{BH}) which arose from (\ref{BA}) is not correct.

This leads us to the role of the multiplicative anomaly. As discussed in \cite{Eliz1,Eliz2} on general grounds we would not expect (\ref{BA}) and (\ref{AL}) to agree, since $\det(A_{1}A_{2})$ will not be the same as $(\det A_{1})(\det A_{2})$ if $A_{1}$ and $A_{2}$ are differential operators. The difference between $\det(A_{1}A_{2})$ and $(\det A_{1})(\det A_{2})$ is called the multiplicative anomaly. Our interpretation of the multiplicative anomaly is not that it is irrelevant to the physics as suggested by Evans \cite{Evans} and Dowker \cite{Dowk}, but rather that it is crucial for obtaining the correct physics. When we factored the differential operator in (\ref{AL}) to obtain (\ref{BA}) we should have included an additional term as discussed in \cite{Eliz1,Eliz2}. The role of this term, which is the multiplicative anomaly, is to ensure that no matter how we handle (\ref{AL}) we end up with the correct result (\ref{AY}). Thus the multiplicative anomaly precisely cancels the first term of (\ref{BH}) resulting in the same answer for the effective action. This does not lead to any new physics as suggested in \cite{Eliz1,Eliz2}, but we do agree with these authors that the multiplicative anomaly is important.

\section{Relativistic Field}
\label{sec3}

As we have already explained, the Feynman functional integral method should lead to a result for the effective action which is in agreement with other methods. Accordingly, we will first use more standard methods to determine what the effective 
action should be, and then see what different functional integral expressions lead to.

\subsection{Thermodynamics and the charge}
\label{sec3.1}

The effective action is related to the Helmholtz free energy, and can therefore be used in a standard way using thermodynamical relations to derive various physical results. Conversely, we can use known results for physical quantities and the thermodynamical relations to tell us the effective action.

Of particular importance to us is the total charge $Q$ which is given by
\begin{equation}
Q = -\frac{1}{\beta}
\left.\frac{\partial\Gamma}{\partial\mu}\right|_{\beta,V}\;.
\label{BM}
\end{equation}
Because a complex scalar field contains particles (of charge $e$) and antiparticles (of charge $-e$) with the Bose-Einstein distribution function, we must have
\begin{equation}
Q = e\sum_{n}\left\{\left[e^{\beta(E_{n}-e\mu)}-1\right]^{-1} - \left[e^{\beta(E_{n}+e\mu)}-1\right]^{-1}\right\}\;.
\label{BN}
\end{equation}
Here $E_{n}$ are the energy eigenvalues. We have
\begin{equation}
E_{n} = \left(\sigma_{n}+m^{2}\right)^{1/2}
\label{BO}
\end{equation}
with $\sigma_{n}$ the eigenvalues of $-\nabla^{2}$. If we now use (\ref{BN}) in (\ref{BM}) and integrate with respect to $\mu$ keeping $\beta$ and $V$ fixed we find
\begin{equation}
\Gamma = \sum_{n}\left\{\ln\left[1-e^{-\beta(E_{n}-e\mu)}\right] + \ln\left[1-e^{-\beta(E_{n}+e\mu)}\right]\right\} + \Gamma_{1}
\label{BP}
\end{equation}
where $\Gamma_{1}$ is independent of $\mu$. We can fix $\Gamma_{1}$ by setting $\mu=0$ and demanding that $\Gamma$ be the result for an uncharged complex scalar field. This fixes $\Gamma_{1}$ to be
\begin{equation}
\Gamma_{1} = \sum_{n}\beta E_{n}\;.
\label{BQ}
\end{equation}
($\Gamma_{1}$ must of course be regularised). $\Gamma_{1}$ has the interpretation of the zero-point energy contribution to $\Gamma$. We end up with the expression
\begin{equation}
\Gamma = \sum_{n}\left\{\beta E_{n} + \ln\left[1-e^{-\beta(E_{n}-e\mu)}\right] + \ln\left[1-e^{-\beta(E_{n}+e\mu)}\right]\right\}
\label{BR}
\end{equation}
which is equivalent to the result found in Ref.~\cite{HW}. It can be expressed as 
\begin{equation}
\Gamma = \Gamma_{+} + \Gamma_{-}
\label{BS}
\end{equation}
where
\begin{equation}
\Gamma_{\pm} = \sum_{n}\left\{\frac{1}{2}\beta(E_{n}\mp e\mu) + \ln\left[1-e^{-\beta(E_{n}\mp e\mu)}\right]\right\}\;.
\label{BT}
\end{equation}
$\Gamma_{+}$ has the interpretation of the particle contribution and $\Gamma_{-}$ the antiparticle contribution. Apart from the different relativistic expression for $E_n$, each of the terms $\Gamma_+$ and $\Gamma_-$ has the same form as we found in the non-relativistic case in (\ref{AX}).

We will call (\ref{BR}) the correct expression for the effective action. It leads, via (\ref{BM}), to the standard result (\ref{BN}) for the charge. In addition, the internal energy
\begin{equation}
U = \left.\frac{\partial\Gamma}{\partial\beta}\right|_{\beta\mu,V}
\label{BU}
\end{equation}
takes the familiar form
\begin{equation}
U = \sum_{n}E_{n}\left\{1 + \left[e^{\beta(E_{n}-e\mu)}-1\right]^{-1} + \left[e^{\beta(E_{n}+e\mu)}-1\right]^{-1}\right\}\;.
\label{BV}
\end{equation}
In the zero temperature limit, the net contribution to $U$ is the zero-point energy contribution, which has its origin in the first term of (\ref{BR}).

\subsection{Partition function and canonical quantisation}
\label{sec3.2}

The standard expression for the partition function is
\begin{equation}
Z = {\rm tr}\ e^{-\beta(H-\mu Q)}
\label{BW}
\end{equation}
with $H$ the Hamiltonian operator and $Q$ the charge operator. The Hamiltonian operator may be expressed as
\begin{equation}
H = \sum_{n}H_{n}
\label{BX}
\end{equation}
with
\begin{equation}
H_{n} = E_{n}\left(a_{n}^{\dagger}a_{n} + \frac{1}{2} + b_{n}^{\dagger}b_{n} + \frac{1}{2}\right)\;.
\label{BY}
\end{equation}
$a_{n}$, $a_{n}^{\dagger}$ are the annihilation and creation operators for particles and $b_{n}$, $b_{n}^{\dagger}$ are those for antiparticles. They satisfy the standard commutation relations for bosons. The charge operator is
\begin{equation}
Q = \sum_{n}Q_{n}
\label{BZ}
\end{equation}
where
\begin{equation}
Q_{n} = e\left(a_{n}^{\dagger}a_{n} - b_{n}^{\dagger}b_{n}\right)
\label{CA}
\end{equation}
It is worth noting that the charge operator has been normal ordered. It is the normal ordered expression which leads to (\ref{BN}).

Because $[H_{n},H_{n'}] = 0 = [Q_{n},Q_{n'}]$ we can write
\begin{equation}
Z = \prod_{n}Z_{n}
\label{CB}
\end{equation}
where
\begin{equation}
Z_{n} = {\rm tr}\ e^{-\beta(H_{n}-\mu Q_{n})}
\label{CC}
\end{equation}
This is a standard manipulation. Finally we can compute $Z_{n}$ by noting that $a_{n}^{\dagger}a_{n}$ is the particle number operator and $b_{n}^{\dagger}b_{n}$ is the antiparticle number operator. Hence we have,
\begin{eqnarray}
Z_{n} & = & \sum_{n'=0}^{\infty}\sum_{n''=0}^{\infty}e^{-\beta E_{n}-\beta(E_{n}-e\mu)n'-\beta(E_{n}+e\mu)n''} \nonumber \\
      & = & e^{-\beta E_{n}}\left[1-e^{-\beta(E_{n}-e\mu)}\right]^{-1} \left[1-e^{-\beta(E_{n}+e\mu)}\right]^{-1}\;.
\label{CD}
\end{eqnarray}
Because the effective action $\Gamma$ is related to the partition function $Z$ by
\begin{equation}
\Gamma = -\ln Z\;,
\label{CE}
\end{equation}
we arrive at
\begin{equation}
\Gamma = \sum_{n}\left\{\beta E_{n} + \ln\left[1-e^{-\beta(E_{n}-e\mu)}\right] + \ln\left[1-e^{-\beta(E_{n}+e\mu)}\right]\right\}\;.
\label{CF}
\end{equation}
This is consistent with the previous result (\ref{BR}).

Thus we can have some faith in the expression (\ref{CF}) which has been arrived at using standard methods which do not involve functional integrals or $\zeta$-function regularisation, as the correct result for the effective action.

\subsection{Functional integral approach}

We now refer back to the result for $\Gamma^{(1)}$ found from the functional integral in (\ref{AG}). The determinant can be defined using $\zeta$-function regularization. We have
\begin{equation}
\Gamma_{A}^{(1)} = -\frac{1}{2}\zeta'_{A}(0) + \frac{1}{2}\zeta_{A}(0)\ln \ell^{4}
\label{CG}
\end{equation}
where
\begin{equation}
\zeta_{A}(s) = \sum_{j=-\infty}^{+\infty}\sum_{n}
\left[\left(\omega_{j}^{2} 
+ \sigma_{n} + m^{2} - e^{2}\mu^{2}\right)^{2} + 
4e^{2}\mu^{2}\omega_{j}^{2}\right]^{-s}\;.
\label{CH}
\end{equation}
(Recall that $\sigma_{n}$ are the eigenvalues of $-\nabla^{2}$.)

Alternatively we could factor the fourth order differential operator in (\ref{AG}) into a product of two second order operators, and define $\Gamma^{(1)}_{B}$ as in (\ref{CI}). If we define
\begin{equation}
\zeta_{B}(s) = \sum_{j=-\infty}^{\infty}\sum_{n}\left[(\omega_{j}+ie\mu)^{2} + \sigma_{n} + m\right]^{-s}
\label{CJ}
\end{equation}
then it is easy to see that
\begin{equation}
\Gamma_{B}^{(1)} = -\zeta'_{B}(0) + \zeta_{B}(0)\ln \ell^{2}
\label{CK}
\end{equation}
is the $\zeta$-function regularised expression. (Both terms in (\ref{CI}) involve the same $\zeta$-function. Because the sum on $j$ in (\ref{CJ}) extends from $-\infty$ to $+\infty$ the sign of $\mu$ in each term is not relevant and each of the two terms gives an identical contribution). 
This result, $\Gamma_{B}^{(1)}$, is the one used in \cite{DJT,KKDJT}, and is equivalent to that used in \cite{Kap,HW,BBD,Kapbook}.

The third result we discussed in Sec.~\ref{sec1} was more complicated and was given in (\ref{CL}). This time we can define two $\zeta$-functions
\begin{equation}
\zeta_{C}^{(1)}(s) = \sum_{j=-\infty}^{+\infty}\sum_{n}\left(\omega_{j}^{2} + \sigma_{n} + m^{2} - e^{2}\mu^{2}\right)^{-s}\;,
\label{CM}
\end{equation}
\begin{equation}
\zeta_{C}^{(2)}(s) = \sum_{j=-\infty}^{+\infty}\sum_{n}\left[\left(\omega_{j}^{2} + \sigma_{n} + m^{2} - e^{2}\mu^{2}\right) + \frac{4e^{2}\mu^{2}\omega_{j}^{2}}{\left(\omega_{j}^{2} + \sigma_{n} + m^{2} - e^{2}\mu^{2}\right)}\right]^{-s}\;.
\label{CN}
\end{equation}
The regularised result for (\ref{CL}) reads
\begin{equation}
\Gamma_{C}^{(1)} = -\frac{1}{2}\zeta_{C}^{(1)}{'}(0) + \frac{1}{2}\zeta_{C}^{(1)}(0)\ln \ell^{2} - \frac{1}{2}\zeta_{C}^{(2)}{'}(0) + \frac{1}{2}\zeta_{C}^{(2)}(0)\ln \ell^{2}\;.
\label{CO}
\end{equation}
The results for $\Gamma_{B}^{(1)}$ and $\Gamma_{C}^{(1)}$ may be regarded as different ways of factoring (\ref{AG}), although $\Gamma_{C}^{(1)}$ has the functional integral interpretation we have mentioned. If we manipulate the formal unregularized expressions for $\Gamma_{A,B,C}^{(1)}$ then they are all identical. However the $\zeta$-function regularised results do not share this equality, as pointed out in \cite{Eliz2}. The only way to decide which, if any, of the expressions for $\Gamma^{(1)}$ is correct is by comparison with a result which does not have any ambiguity. We have found in Secs.~\ref{sec3.1} and  \ref{sec3.2} that the standard result (\ref{CF}) holds. (Of course we have only written down three possible expressions for $\Gamma$, and there are many other ways to evaluate the functional integral).

We turn first to $\Gamma_{B}^{(1)}$ because it is the easiest to evaluate, and as we shall show leads to a result in agreement with (\ref{CF}). We will first show this formally using the result of the Appendix, specifically (\ref{App8}). To do 
this we need to know the behaviour of the energy $\zeta$-function $E(2s)$ near $s=0$. We have
\begin{equation}
E(2s) = \sum_{n}E_{n}^{1-2s} = \sum_{n}\left(\sigma_{n} + m^{2}\right)^{1/2-s}\;.
\label{CP}
\end{equation}
Although it is possible to proceed generally, without knowing the eigenvalues $\sigma_{n}$ of $-\nabla^{2}$ explicitly, we will for simplicity specialise to flat space in the infinite volume limit. In this case (\ref{CP}) becomes
\begin{eqnarray}
E(2s) & = & V\int\frac{d^{3}k}{(2\pi)^{3}}\left(k^{2} + m^{2}\right)^{1/2-s} \nonumber \\
 & = & \frac{V}{8\pi\sqrt{\pi}}\frac{\Gamma(s-2)}{\Gamma(s-1/2)}\left(m^{2}\right)^{2-s}\;.
\label{CQ}
\end{eqnarray}
The first term of (\ref{App8}) therefore involves
\begin{equation}
\frac{\beta}{2\sqrt{\pi}}\frac{\Gamma(s-1/2)}{\Gamma(s)}E(2s) = \frac{\beta V}{16\pi^{2}}\frac{(m^{2})^{2-s}}{(s-1)(s-2)}
\label{CR}
\end{equation}
\begin{equation}
= \frac{\beta V}{32\pi^{2}}m^{4}\left[1 + \frac{3}{2}s - s\ln m^{2} + \cdots\right]
\label{CS}
\end{equation}
when expanded about $s=0$. We can use (\ref{App8}) to conclude that
\begin{eqnarray}
\zeta_{B}(0) & = & \frac{\beta V}{32\pi^{2}}m^{4}\;, \nonumber \\
\zeta_{B}'(0) & = & \frac{\beta V}{32\pi^{2}}m^{4}\left(\frac{3}{2}-\ln m^{2}\right) \nonumber \\
& & \: - \sum_{n}\left\{\ln\left[1-e^{-\beta(E_{n}-e\mu)}\right] + \ln\left[1-e^{-\beta(E_{n}-e\mu)}\right]\right\}\;. \nonumber
\end{eqnarray}
Thus
\begin{eqnarray}
\Gamma_{B}^{(1)} & = & \sum_{n}\left\{\ln\left[1-e^{-\beta(E_{n}-e\mu)}\right] + \ln\left[1-e^{-\beta(E_{n}-e\mu)}\right]\right\} \nonumber \\
 & & \:- \frac{\beta V}{32\pi^{2}}m^{4}\left(\frac{3}{2}-\ln(m^{2}\ell^{2})\right)\;.
\label{CT}
\end{eqnarray}
The last term in (\ref{CT}), which is independent of the chemical potential $\mu$, contains the contribution of the zero-point energy after regularisation.

We now turn to the high temperature expansion of $\Gamma_{B}^{(1)}$. Haber and Weldon \cite{HW} obtained a result which ignored the zero-point energy, and we will show (following ~\cite{DJT}) how to obtain their result directly from the generalised $\zeta$-function. Separating off the $j=0$ term in (\ref{CJ}) we have
\begin{equation}
\zeta_{B}(s) = \tilde{\zeta}_{B}(s) + F_{+}(s) + F_{-}(s)
\label{CU}
\end{equation}
where
\begin{equation}
\zeta_{B}(s) = \sum_{n}\left(\sigma_{n} + m^{2} -
 e^{2}\mu^{2}\right)^{-s}\;,
\label{CV}
\end{equation}
\begin{equation}
F_{\pm}(s) = \sum_{j=1}^{\infty}\sum_{n}\left[(\omega_{j} \pm ie\mu)^{2} + \sigma_{n} + m^{2}\right]^{-s}\;.
\label{CW}
\end{equation}
Specialising to flat space, replacing $\sigma_{n} \rightarrow k^{2}$ and $\sum_{n} \rightarrow V\int\frac{d^{3}k}{(2\pi)^{3}}$ we find
\begin{equation}
\tilde{\zeta}_{B}(s) = \frac{V}{8\pi\sqrt{\pi}}\frac{\Gamma(s-3/2)}{\Gamma(s)}\left(m^{2} - e^{2}\mu^{2}\right)^{3/2-s}\;,
\label{CX}
\end{equation}
\begin{equation}
F_{\pm}(s) = \frac{V}{8\pi\sqrt{\pi}}\frac{\Gamma(s-3/2)}{\Gamma(s)}\left(\frac{2\pi}{\beta}\right)^{3-2s} \sum_{j=1}^{\infty}\left[(j \pm ie\bar{\mu})^{2} + \bar{m}^{2}\right]^{3/2-s}\;,
\label{CY}
\end{equation}
with $\bar{\mu}=\beta\mu/2\pi$ and $\bar{m}=\beta m/2\pi$. The binomial expansion may be used to evaluate the leading terms of $F_{\pm}(s)$ in the high temperature limit. After a bit of calculation we find
\begin{equation}
F_{+}(0) + F_{-}(0) = \frac{\beta V}{32\pi^{2}}m^{4}\;,
\label{CZ}
\end{equation}
(which is an exact result), and
\begin{eqnarray}
\frac{d}{ds}\left.\left(F_{+}(s)+F_{-}(s)\right)\right|_{s=0} & = & \frac{4\pi^{2}V}{3\beta^{3}}\left(\frac{1}{60} + \frac{\beta^{2}e^{2}\mu^{2}}{8\pi^{2}} - \frac{\beta^{4}e^{4}\mu^{4}}{32\pi^{2}}\right) \nonumber \\
 & & \:\:+\frac{V}{2\beta}m^{2}\left(-\frac{1}{6} + \frac{\beta^{2}e^{2}\mu^{2}}{4\pi^{2}}\right) \nonumber \\
 & & \:\:+\frac{\beta V}{32\pi^{2}}m^{4}\left(2\gamma + 2\ln\frac{\beta}{4\pi}\right) + \cdots\;,
\label{DA}
\end{eqnarray}
where $\gamma$ is the Euler-Mascheroni constant. The expansion for $\Gamma_{B}^{(1)}$ becomes
\begin{eqnarray}
\Gamma_{B}^{(1)} & = & -\frac{V}{6\pi}(m^{2}-e^{2}\mu^{2})^{3/2} - \frac{\pi^{2}V}{45\beta^{3}} + \frac{V}{12\beta}(m^{2}-2e^{2}\mu^{2}) \nonumber \\
 & & \: -\frac{\beta V}{16\pi^{2}}m^{4}\left(\gamma + \ln\frac{\beta}{4\pi \ell}\right) + \frac{\beta V}{24\pi^{2}}e^{2}\mu^{2}(e^{2}\mu^{2}-3m^{2}) + \cdots\;.
\label{DB}
\end{eqnarray}
Removing the zero-point energy term using (\ref{CT}) shows complete agreement between this result and that of Haber and Weldon \cite{HW}. We can safely conclude that the correct physics is contained in the expression $\Gamma_{B}^{(1)}$.

We now turn to the other two expressions $\Gamma_{A}^{(1)}$ and $\Gamma_{C}^{(1)}$. We have not found such an elegant way to analyse the $\zeta$-functions in these two cases as that presented in the Appendix. Instead we will content ourselves with the high temperature limit only and compare with (\ref{DB}). We will show that different results are obtained in these two cases.

We take $\zeta_{A}(s)$ in (\ref{CH}) and expand in powers of $\mu$, keeping terms up to order $\mu^{4}$. If we define
\begin{equation}
G(z,k) = \sum_{j=-\infty}^{+\infty}\sum_{n}\omega_{j}^{2k}\left(\omega_{j}^{2} + \sigma_{n} + m^{2} - e^{2}\mu^{2}\right)^{-z}
\label{DC}
\end{equation}
it is easy to show that
\begin{equation}
\zeta_{A}(s) = G(2s,0) - 4e^{2}\mu^{2}sG(2s+2,1) + 8e^4\mu^{4}s(s+1)G(2s+4,2) + \cdots\;.
\label{DD}
\end{equation}
Taking the case of flat space in the large box limit we find
\begin{equation}
G(z,k) = \frac{V}{(4\pi)^{3/2}}\frac{\Gamma(z-3/2)}{\Gamma(z)}
\sum_{j=-\infty}^{+\infty}\omega_{j}^{2k}\left(\omega_{j}^{2} + m^{2} - 
e^{2}\mu^{2}\right)^{3/2-z}\;.
\label{DE}
\end{equation}
For $k\geq 1$, we can expand $G(z,k)$ to find (noting that the $j=0$ term in the sum makes no contribution for $k\geq 1$)
\begin{eqnarray}
G(z,k) & = & \frac{V}{4\pi\sqrt{\pi}}\frac{\Gamma(z-3/2)}{\Gamma(z)}\left\{\left(\frac{2\pi}{\beta}\right)^{3+2k-2z}\zeta_{R}(2z-3-2k)\right. \nonumber \\
 & & \left.\: +\left(\frac{3}{2}-z\right)(m^{2}-e^{2}\mu^{2})\left(
\frac{2\pi}{\beta}\right)^{1+2k-2z}\zeta_{R}(2z-1-2k)\right. \nonumber \\
 & & \left.\: +\frac{1}{2}\left(\frac{3}{2}-z\right)
(m^{2}-e^{2}\mu^{2})^{2}
\left(\frac{2\pi}{\beta}\right)^{2k-1-2z}\zeta_{R}(2z+1-2k)\right. 
\nonumber \\
 & & \left.\: + \cdots\rule{0mm}{7mm}\right\}\;.
\label{DF}
\end{eqnarray}
Here $\zeta_{R}(\alpha)$ denotes the Riemann $\zeta$-function. For $k=0$ the $j=0$ term does make a contribution and we find
\begin{eqnarray}
G(z,0) & = & \frac{V}{8\pi\sqrt{\pi}}\frac{\Gamma(z-3/2)}{\Gamma(z)}(m^{2}-e^{2}\mu^{2})^{3/2-z} \nonumber \\
 & & \: + \frac{V}{4\pi\sqrt{\pi}}\frac{\Gamma(z-3/2)}{\Gamma(z)}\left\{\left(\frac{2\pi}{\beta}\right)^{3-2z}\zeta_{R}(2z-3)\right. \nonumber \\
 & & \left.\: + \left(\frac{3}{2}-z\right)(m^{2}-e^{2}\mu{2})\left(\frac{2\pi}{\beta}\right)^{1-2z}\zeta_{R}(2z-1)\right. \nonumber \\
 & & \left.\: + \frac{1}{2}\left(\frac{3}{2}-z\right)\left(\frac{1}{2}-z\right)\left(\frac{2\pi}{\beta}\right)^{-1-2z}(m^{2}-e^{2}\mu^{2})\zeta_{R}(2z+1)\right. \nonumber \\
 & & \left.\: + \cdots\rule{0mm}{7mm}\right\}\;.
\label{DG}
\end{eqnarray}
These results are sufficient to show that
\begin{equation}
\zeta_{A}(0) = \frac{\beta V}{32\pi^{2}}m^{4}\;,
\label{DH}
\end{equation}
\begin{eqnarray}
\zeta'_{A}(0) & = & \frac{V}{3\pi}(m^{2}-e^{2}\mu^{2})^{3/2} + \frac{2\pi^{2}V}{45\beta^{3}} - \frac{V}{6\beta}(m^{2}-2e^{2}\mu^{2}) \nonumber \\
 & & \: + \frac{\beta V}{8\pi^{2}}m^{4}\left(\gamma+\ln\frac{\beta}{4\pi}\right) + \frac{\beta V}{8\pi^{2}}e^{2}\mu^{2}\left(m^{2}-\frac{1}{3}e^{2}\mu^{2}\right) + 
\cdots\;.
\label{DI}
\end{eqnarray}
Used in (\ref{CG}) we find
\begin{eqnarray}
\Gamma_{A}^{(1)} & = & -\frac{V}{6\pi}(m^{2}-e^{2}\mu^{2})^{3/2} - \frac{\pi^{2}V}{45\beta^{3}} + \frac{V}{12\beta}(m^{2}-2e^{2}\mu^{2}) \nonumber \\
 & & \: -\frac{\beta V}{16\pi^{2}}m^{4}\left(\gamma+\ln\frac{\beta}{4\pi \ell}\right) + \frac{\beta V}{48\pi^{2}}e^{2}\mu^{2}(e^{2}\mu^{2}-3m^{2}) + \cdots\;.
\label{DJ}
\end{eqnarray}
Comparison with (\ref{DB}) shows that all terms are the same apart from the last one. In fact the difference between $\Gamma_{A}^{(1)}$ and $\Gamma_{B}^{(1)}$ corresponds precisely to the multiplicative anomaly computed in \cite{Eliz2}. We will return to this at the end of the section.

Finally we will examine the high temperature limit of $\Gamma_{C}^{(1)}$. Again a straightforward binomial expansion can be used. Leaving out the details, we find
\begin{eqnarray}
\zeta_{C}^{(1)}(0) & = & \frac{\beta V}{32\pi^{2}}(m^{2}-
e^{2}\mu^{2})^{2}\;, \nonumber \\
\zeta_{C}^{(1)\prime}(0) & = & \frac{V}{3\pi}(m^{2}-e^{2}\mu{2})^{3/2} + \frac{\pi^{2}V}{45\beta^{3}} - \frac{V}{12\beta}(m^{2}-e^{2}\mu^{2}) \nonumber \\
 & & \: +\frac{\beta V}{16\pi^{2}}(m^{2}-e^{2}\mu^{2})^{2}\left(\gamma+
\ln\frac{\beta}{4\pi}\right) + \cdots\;, \nonumber \\
\zeta_{C}^{(2)}(0) & = & \frac{\beta V}{16\pi^{2}}e^{2}\mu^{2}(2m^{2}-e^{2}\mu^{2})\;, \nonumber \\
\zeta_{C}^{(2)\prime}(0) & = & \zeta_{C}^{(1)\prime}(0) + \frac{V}{6\beta}e^{2}\mu^{2} \nonumber \\
 & & \: +\frac{\beta V}{4\pi^{2}}e^{2}\mu^{2}(m^{2}-e^{2}\mu^{2})\left(\gamma+\frac{1}{2}
+\ln\frac{\beta}{4\pi}\right) \nonumber \\
 & & \: +\frac{\beta V}{8\pi^{2}}e^{4}\mu^{4}\left(\gamma+\frac{11}{12}+
\ln\frac{\beta}{4\pi}\right) + \cdots\;. \nonumber
\end{eqnarray}
These results lead to
\begin{eqnarray}
\Gamma_{C}^{(1)} & = & -\frac{V}{6\pi}(m^{2}-e^{2}\mu^{2})^{3/2} - \frac{\pi^{2} V}{45\beta^{3}} + \frac{V}{12\beta}(m^{2}-2e^{2}\mu^{2}) \nonumber \\
 & & \: -\frac{\beta V}{16\pi^{2}}m^{4}\left(\gamma+\ln\frac{\beta}{4\pi \ell}\right) - \frac{\beta V}{16\pi^{2}}e^{2}\mu^{2}\left(m^{2}-\frac{1}{12}e^{2}\mu^{2}\right) + \cdots\;.
\label{DK}
\end{eqnarray}
Again it is only the final term which differs from the result (\ref{DB}).

We are in the situation that we have three different ways of evaluating the formal expression for the effective action. Only one of these expressions, $\Gamma_{B}^{(1)}$ corresponds to a result found using canonical methods. The difference between the three results for the effective action cannot be due to the fact that we have not renormalized $\Gamma$. The only renormalization ambiguity resides in our choice of the renormalization length $\ell$. Rescaling $\ell$ in any of the expressions we have found only alters the effective action by a term proportional to $\beta Vm^{4}$, which does not involve $\mu$. The only way to decide which of the expressions we have found is correct is by comparison with physical results as we have done. (Alternatively a more careful definition of the functional integral might settle the issue, but it must lead to $\Gamma_{B}^{(1)}$ if results of standard statistical mechanics are to be correct). To emphasise this point as clearly as possible, we can compute the charge in the high temperature limit using (\ref{BM}). Because the $\mu$-dependence in $\Gamma_{A}^{(1)},\Gamma_{B}^{(1)}$ and $\Gamma_{C}^{(1)}$ are all different, we will obtain three different results for the charge. Only one of these results can correspond to 
that found from a direct high temperature expansion of (\ref{BN}), and this comes from $\Gamma_{B}^{(1)}$.

At this stage we return to the multiplicative anomaly. Although we claim that it does not lead to any new physics, we do not agree with the authors of Refs.~\cite{Dowk,Evans} that it is of no importance. Rather we support the view that it is extremely important. Having settled on the correct expression for the effective action $\Gamma_{B}^{(1)}$ in (\ref{CI}), the multiplicative anomaly is what ensures that if we combine the two operators to obtain (\ref{AG}) the correct result will be obtained. The reason is that it is necessary to add 
on the anomaly term when the operators are combined as found in \cite{Eliz1,Eliz2} and this leads to the same result as found from $\Gamma_{B}^{(1)}$. A similar comment applies to $\Gamma_{C}^{(1)}$ with a different anomaly because of the different factorization. The multiplicative anomaly guarantees that formal manipulations of different factorizations of the effective action all lead to the same physics.

\section{Discussion and conclusions}

We have argued in the preceding sections that to determine the validity of the inequivalent but formally identical expressions for the effective action obtained from the functional integral, physical considerations are of greater importance than mathematical ones. The functional integral method is merely one way of calculation, and obviously all valid methods of finding the effective action must lead to the same physical conclusions. We studied both the relativistic and non-relativistic scalar fields. In 
the non-relativistic case two results for  the
effective action, which are equivalent at the formal level, were evaluated, and the one which agreed with the results of standard thermodynamics was identified. In the relativistic case we evaluated three possible expressions for the effective action. The
 correct expression was identified both from looking at the charge, and from a direct evaluation of the partition function using canonical methods. Once the correct expression had been found we emphasized the crucial role of the multiplicative anomaly in 
maintaining equality between the correct expression and other formally equivalent expressions.

We have restricted our attention to non-interacting scalar field theories here. The role of interactions complicates the details, but does not lead to any
differences of substance. At zero temperature the presence of a quartic self-interaction for the scalar field leads to a term in the multiplicative anomaly proportional to the interaction as found in \cite{Eliz1}. However in this case the anomaly term is 
of no physical significance, in contrast to the view
taken in \cite{Eliz1}, because the effective action (or potential) has not been renormalized. Once a renormalization condition has been imposed the anomaly is absorbed by the counterterms and the usual effective potential is obtained. At finite temperature the situation is slightly more complicated \cite{Eliz2}, but the approach we have outlined above settles the issue in favour of the standard expression for the effective action as used in \cite{Kap,HW,BBD,DJT,KKDJT} for  example.

It is also worth commenting on the theory used in two recent criticisms \cite{Dowk,Evans} of the work in \cite{Eliz1,Eliz2}. This model consists of two non-interacting scalar fields of different masses. The effective action can be expressed in 
two ways~:
\begin{eqnarray}
\Gamma_1 & = & -\frac{1}{2}\ln\det\ell^4(-\Box_x + m_1^2)(-\Box_x + m_2^2)\;, \\
\Gamma_2 & = & -\frac{1}{2}\ln\det\ell^2(-\Box_x + m_1^2) \nonumber\\
 & &\quad -\frac{1}{2}\ln\det\ell^2(-\Box_x + m_2^2)\;.
\label{DL}
\end{eqnarray}
These expressions are formally equal. As shown in \cite{Eliz1} if $\zeta$-function
regularization is used there is a multiplicative anomaly present so that an explicit evaluation leads to $\Gamma_1\ne\Gamma_2$. However this theory is too simple to settle any issues about the role of the multiplicative anomaly. The difference between the
 two expressions $\Gamma_1$ and $\Gamma_2$ is proportional to $\beta V(m_1^2-m_2^2)^2$. This is a constant term in the effective potential which has no physical significance. As for the self-interacting theory at zero temperature, the anomaly is unimportant once a renormalized result is considered by adopting a renormalization condition, in this case on the vacuum energy.

In conclusion, we support the authors of \cite{Eliz1,Eliz2} that the multiplicative anomaly is important, but we do not agree that it contains any new physical consequences. Rather the multiplicative anomaly is needed to explain the equality of 
formally identical expressions for the effective action arising from the functional integral.

\acknowledgments

J.~J.~M.-S. would like to thank EPSRC for grant 97304113.

\appendix
\section*{The generalised $\zeta$-function}

We have the general definition
\begin{equation}
\zeta(s) = \sum_{j=-\infty}^{+\infty}\sum_{n}\left[(\omega_{j} + i\mu)^{2} + E_{n}^{2}\right]^{-s}\;.
\label{App1}
\end{equation}
This will be an analytic function of $s$ in some region of the complex plane and the objective is to analytically continue it to a neighbourhood of $s=0$ and find $\zeta(0)$ and $\zeta'(0)$. There are many ways to do this, and we will outline one way here.

The order of summations is irrelevant in the region of the complex $s$-plane where (\ref{App1}) converges. We will perform the sum over $j$ first. By making use of the summation formula
\begin{equation}
\sum_{j=-\infty}^{+\infty}f(j) = \int_{-\infty}^{+\infty}f(j)dj + \int_{-\infty + i\epsilon}^{\infty + i\epsilon}dz\left(e^{-2\pi iz}-1\right)^{-1}\left[f(z)+f(-z)\right]
\label{App2}
\end{equation}
we obtain
\begin{eqnarray}
\zeta(s) & = &  \frac{\beta}{2\sqrt{\pi}}\frac{\Gamma(s-1/2)}{\Gamma(s)}\sum_{n}E_{n}^{1-2s} + \sum_{n}\int_{-\infty + i\epsilon}^{\infty + i\epsilon}dz\left(e^{-2\pi iz}-1\right)^{-1} \nonumber \\
 & & \:\times\;\left\{\left[\left(\frac{2\pi z}{\beta}+i\mu\right)^{2}+E_{n}^{2}\right]^{-s} + \left[\left(\frac{2\pi z}{\beta}-i\mu\right)^{2}+E_{n}^{2}\right]^{-s}\right\}\;.
\label{App3}
\end{eqnarray}
In arriving at the first term we have made use of the definition of the $\Gamma$-function. If we define an energy $\zeta$-function by
\begin{equation}
E(\alpha) = \sum_{n}E_{n}^{1-\alpha}
\label{App4}
\end{equation}
which will be analytic for $\Re(\alpha)$ large enough, we can write
\begin{eqnarray}
\zeta(s) & = & \frac{\beta}{2\sqrt{\pi}}\frac{\Gamma(s-1/2)}{\Gamma(s)}E(2s) + \sum_{n}\int_{-\infty + i\epsilon}^{\infty + i\epsilon}dz\left(e^{-2\pi iz}-1\right)^{-1} \nonumber \\
 & & \times\;\left\{\left[\left(\frac{2\pi z}{\beta}+i\mu\right)^{2}+E_{n}^{2}\right]^{-s} + \left[\left(\frac{2\pi z}{\beta}-i\mu\right)^{2}+E_{n}^{2}\right]^{-s}\right\}\;.
\label{App5}
\end{eqnarray}

In order to obtain the analytic continuation of $\zeta(s)$ to $s=0$ we must modify (\ref{App5}) since the second term of (\ref{App5}) diverges at $s=0$ as it stands. The integrand has branch points at $2\pi z/\beta = i(E_{n}\pm\mu)$ in the upper half plane. By taking branch cuts along the imaginary axis, and deforming the contour around the branch cuts it is straightforward to show that
\begin{eqnarray}
\zeta(s) & = & \frac{\beta}{2\sqrt{\pi}}\frac{\Gamma(s-1/2)}{\Gamma(s)}E(2s) \nonumber \\
 & & \: + 2\sin(\pi s)\sum_{n}\left\{\int_{\frac{\beta(E_{n}-\mu)}{2\pi}}^{\infty}dx
\left(e^{2\pi x} - 1\right)^{-1}\left[\left(\frac{2\pi}{\beta}x+\mu\right)^{2} - E_{n}^{2}\right]^{-s}\right. \nonumber \\
 & & \: \left.+ \int_{\frac{\beta(E_{n}+\mu)}{2\pi}}^{\infty}dx\left(e^{2\pi x} - 1\right)^{-1}\left[\left(\frac{2\pi}{\beta}x-\mu\right)^{2} - E_{n}^{2}\right]^{-s}\right\}\;.
\label{App7}
\end{eqnarray}
(An equivalent analytic continuation was found earlier by Ford \cite{Ford}). By expanding the second term in (\ref{App7}) about $s=0$ it is easy to show that
\begin{eqnarray}
\zeta(s) & = & \frac{\beta}{2\sqrt{\pi}}\frac{\Gamma(s-1/2)}{\Gamma(s)}E(2s) \nonumber \\
 & & - s\sum_{n}\left\{\ln\left(1-e^{-\beta(E_{n}-\mu)}\right) + \ln\left(1-e^{-\beta(E_{n}+\mu)}\right)\right\} + \cdots
\label{App8}
\end{eqnarray}
where terms of order $s^{2}$ and higher have been dropped. The details of expanding the first term depend on the spectrum $E_{n}$. Another way of obtaining (\ref{App8}) is given in the appendix of \cite{Smith}.

\end{document}